\documentstyle[preprint,aps]{revtex}

\input epsf         
\epsfverbosetrue    

\begin{document}

\draft

\title{The Surface Region of Superfluid $^4$He as a Dilute Bose-Condensed Gas}

\author{A. Griffin}
\address{Department of Physics, University of Toronto,
         Toronto, Ontario, Canada, M5S 1A7} 

\author{S. Stringari}
\address{Dipartimento di Fisica and INFM, Universit\`{a} di Trento,
I-38050 Povo, Italy}


\maketitle

\begin{abstract} 
In the low-density surface region of superfluid
$^4$He, the atoms are far apart and collisions can be ignored.
The only effect of the interactions is
from the long-range attractive Hartree potential produced by the distant
high-density bulk liquid. As a result, at $T=0$, all the atoms occupy the same
single-particle state in the low-density tail.
Striking numerical evidence for this 100\% surface BEC was given by
Pandharipande and coworkers in 1988.
We derive a generalized Gross-Pitaevskii equation
for the inhomogeneous condensate wave function
$\Phi(z)$ in the low-density region
valid at all temperatures.
The overall amplitude of $\Phi(z)$ is fixed by the bulk liquid, which
ensures that it vanishes everywhere at the bulk transition
temperature.
\end{abstract}

\vskip 0.1 true in


\narrowtext

\newpage

\section{INTRODUCTION}

Our starting point \cite{ref1} is very simple, namely that
at the free surface of superfluid $^4$He, the local density of
atoms $n({\bf r})$ becomes small. Since the
system is like a dilute inhomogeneous Bose gas of $^4$He atoms, the
effect of inter-atomic collisions can be ignored. The dominant
effect of the interactions is from the long-range
attractive Hartree potential produced by
the bulk liquid. The final result is that when $n({\bf r})$
is less than 10\% of the bulk density $n_B$, at $T=0$
one has almost complete Bose-Einstein condensation (see Fig.~1 of
Ref.~\cite{ref1}).
That is to say, all the $^4$He atoms in the
low-density tail are in the identical single-particle state.
This unusual surface Bose condensate should exhibit striking features
when probed in an appropriate way (this ``surface'' condensate is not a
2D phenomenon, since the low-density
region is spread over a large distance from the free surface).

A key difference from the recently observed trapped Bose-condensed atomic
gases is that through the long-range attractive tail of the
interatomic potential, the
bulk liquid determines the surface density profile and the
amplitude of the Bose order parameter. The surface condensate is
``driven'' by the bulk liquid, which ensures that it vanishes in
both the surface and bulk regions at the same bulk transition temperature.
In this sense, the surface Bose condensate might be viewed as a
realization of the ``atom laser'' which is being
currently discussed in connection with trapped atomic gases.

\section{CONDENSATE WAVEFUNCTION IN SURFACE REGION}

Separating out the condensate contribution in the quantum field
operators in the usual way, we have
\begin{eqnarray}
\hat{\psi}({\bf r})=\Phi({\bf r})+\tilde{\psi}({\bf r})~~~~\nonumber\\
\hat{\psi}^\dagger({\bf r})=\Phi^*({\bf r})+\tilde{\psi}^\dagger({\bf r}),
\label{eq1}
\end{eqnarray}
where the condensate wavefunction $\Phi({\bf r})\equiv
\langle\hat{\psi}({\bf r})\rangle$
is the order parameter of a Bose-condensed system.
The total density is given by
\begin{equation}
\langle\hat{\psi}^\dagger{\bf r})\hat{\psi}({\bf r})\rangle\equiv
n({\bf r})=n_c({\bf r})+\tilde{n}({\bf r}),
\label{eq2}
\end{equation}
where $n_c({\bf r})=|\Phi({\bf r})|^2$ is the local condensate and
$\tilde{n}({\bf r})\equiv
\langle\tilde{\psi}^\dagger{\bf r})\tilde{\psi}({\bf r})\rangle$ is
the local density of $^4$He atoms out of the condensate.
The equation of motion for $\hat{\psi}({\bf r})$ leads to an
exact equation for $\Phi({\bf r})$ given by \cite{ref3,ref1}
\begin{equation}
  \left [ -\frac{\hbar^2\nabla^2}{2m}-\mu\right ]\Phi({\bf r})
+\int
  d{\bf r}^\prime v({\bf r} - {\bf r}^\prime)\langle
  \hat{n}({\bf r}^\prime)\hat\psi({\bf
r})\rangle=0\ ,
\label{eq3}
\end{equation}
where $v({\bf r}-{\bf r}^{\prime})$ is the bare He-He interaction.

In a uniform Bose fluid, (\ref{eq3})
determines the bulk
chemical potential $\mu$ in terms of $\Phi$. In a
system with a free surface in thermal equilibrium, one can find the
asymptotic spatial dependence of $\Phi({\bf r})$ where
the density $n({\bf r})$ is small, since then (\ref{eq3}) reduces to
\begin{equation}
  \left[-\frac{\hbar^2\nabla^2}{2m}-\mu +v_H({\bf r}) \right] \Phi({\bf r})=0,
\label{eq4}
\end{equation}
where
\begin{equation}
  v_H({\bf r}) =  \int^\prime d{\bf r}^\prime v({\bf r} -
  {\bf r}^\prime) n({\bf r}^\prime)
\label{eq5}
\end{equation}
is the Hartree potential at a point in
the low-density region
due to contributions from ${\bf r}^\prime$ in the bulk region.
Since $|{\bf r}-{\bf r}^\prime|$
is large, only the long-range attractive part of the He-He potential
is relevant in (\ref{eq5}).
Contributions when ${\bf r}^\prime$ is close to ${\bf r}$
are negligible since the density is so small
(for further discussion, see \cite{ref1}).

Far enough away from a planer surface centered at $z=0$,
$v_H({\bf r})$ in (\ref{eq5})
does not depend on the detailed form of the surface profile
$n({\bf r})$. To leading order, we have
$v_{H}(z) = -n_{B} \alpha/z^{3}$, with $\alpha$ being determined
by $v({\bf r})$.
This allows us to solve (\ref{eq4}) for the asymptotic spatial dependence
\begin{equation}
\Phi (z)=Ce^{-(Az+B/z^2)},
\label{eq6}
\end{equation}
where $A = \sqrt{\frac{2m\vert\mu\vert}{\hbar^2}} \simeq 1\AA^{-1}$
  and $B = m \rho _{B} \alpha/2 \hbar^{2} A$.
While we claim this gives the exact asymptotic behaviour, we
cannot calculate the amplitude $C$ in (\ref{eq6})
since this is determined by the bulk liquid.

The preceeding analysis leading to (\ref{eq6})
is valid at all temperatures below the bulk
superfluid transition. We note that  (\ref{eq4}) relates $\Phi({\bf r})$ in the
low-density region to the density profile $n({\bf r})$
in the bulk region, for which
we do not have a microscopic theory.
However, $T=0$ variational Monte Carlo calculations \cite{ref2} carried out
some years ago show that in the
low-density region, $n({\bf r})$ approaches
$n_{c}({\bf r})=|\Phi({\bf r})|^2$ and becomes identical with it
when $n({\bf r})~\alt~ 0.1n_B$.
Thus, in this region (see Fig.~1 of Ref.~\cite{ref1}),
we have complete 100\% BEC, with the local density
given by the square of (\ref{eq6}).

While (\ref{eq6}) is still valid,
the amplitude $C$ will decrease towards zero as we approach the
bulk superfluid transition due to depletion of the condensate.
At finite temperatures, the non-condensate local density $\tilde{n}({\bf r})$
will asymptotically approach the finite (but very
small) uniform vapour density
appropriate to a gas-liquid system in thermal equilibrium (This
vapour is always a classical dilute gas since $|\mu|\gg k_B T$).

As one gets near the interface, of course, $n({\bf r})$ will
increasingly deviate from $n_c({\bf r})$. We cannot solve (\ref{eq3}) in
this region. However, we call attention to the fact that the
absolute value of the condensate density
is expected to peak at the interface \cite{ref1,ref2}.
According to the numerical results of Ref.~\cite{ref2},
one finds $n_c({\bf r})\simeq 0.2n_B$ when $n({\bf r})\simeq 0.5n_B$.
This peak in the condensate density at the interface should be
looked for experimentally.

\section{CONCLUSIONS}

We have argued \cite{ref1,ref2} that at $T=0$,
all the atoms are in the same single-particle quantum state in
the low-density tail of superfluid $^4$He.
This inhomogeneous Bose-condensed gas is ``controlled''
by the bulk liquid through the long-range attractive tail of the
He-He interaction potential.

A time-dependent perturbation $v_{ex}({\bf r},t)$
of the bulk liquid will drive this surface condensate coherently.
Using the same physical arguments given in deriving (\ref{eq4}),
the equation of motion for
$\Phi({\bf r},t)\equiv \langle\hat{\psi}({\bf r})\rangle_t$ is

\begin{equation}
i\hbar{\partial \Phi({\bf r},t)\over\partial t}=
\left[-\frac{\hbar^2\nabla^2}{2m} +v_{ex}({\bf r},t)
+v_{H}({\bf r},t)\right ] \Phi({\bf r},t)
\label{eq7}
\end{equation}
with the time-dependent Hartree potential given by
\begin{equation}
v_{H}({\bf r},t)=\int^\prime d{\bf r}^\prime
v({\bf r-r^\prime})\langle\hat{n}({\bf r}^\prime)\rangle_t.
\label{eq8}
\end{equation}
As with (\ref{eq4}), (\ref{eq7}) is only valid for ${\bf r}$
in the low-density region and (\ref{eq8})
only includes the contributions from ${\bf r}^\prime$
in the high-density bulk region.
Experiments to probe (and utilize) this surface Bose
condensate are needed, especially in connection with the
unique coherent features it should exhibit.

\acknowledgements
The work of A.G. was supported by a research grant from NSERC of Canada.

\end{document}